\documentclass[reprint, superscriptaddress, showkeys, twocolumn]{revtex4-1}
\usepackage{graphicx}
\usepackage{amsmath}
\usepackage{natbib}
\begin{document}
\title{Effects of Residual Stress on Static and Dynamic Characteristics of an Electrostatically Actuated Nanobeam}
\author{A. Bhushan}
\affiliation{Department of Mechanical Engineering, Indian Institute of Technology Bombay, Mumbai - 400076, India.}
\author{M. M. Inamdar}
\affiliation{Department of Civil Engineering, Indian Institute of Technology Bombay, Mumbai - 400076, India.}
\author{D. N. Pawaskar}
\thanks{Corresponding author}
\email[Email address: ]{pawaskar@iitb.ac.in}
\affiliation{Department of Mechanical Engineering, Indian Institute of Technology Bombay, Mumbai - 400076, India.}

\begin{abstract}
Electrostatically actuated nanotubes and nanowires have many promising applications as nano-switches, ultra sensitive sensors and signal processing elements. These devices can be modelled as slender beams with circular cross-section. In this paper, effects of residual stress on static and dynamic characteristics of a cylindrical nanobeam are presented. Galerkin based multi-modal reduced order model technique has been used to solve the governing differential equation. The equation has also been solved numerically by collocation method for verification of the results. The effects of higher modes in reduced order modelling have been investigated. The analysis shows that the residual stress significantly influence the static and dynamic characteristics of the nanobeam. In particular, the results show that resonating frequency tunability increases significantly under the presence of compressive residual stress. In addition, an inherent resonating frequency stability mechanism under temperature variation in nanobeam resonators has also been explored.
\end{abstract}
\maketitle
\section{Introduction}
Nanoelectromechanical systems (NEMS) is a vibrant field of research. Many NEMS devices have been developed as ultrasensitive sensors \cite{yang2006, stamfer2006}, nano-switches \cite{ subramanian2009} and  RF communication elements \cite{rutherglen2009}, based on electrostatically actuated nanotubes and nanowires. In electrostatic actuation, a nanotube or nanowire is in capacitive arrangement with a plate electrode and bias voltage between them creates actuation force. DC voltage is responsible for static deflection while AC voltage creates oscillatory motion. Efforts have been going on to understand the various static and dynamic characteristics of nanotubes and nanowires based devices \cite{dequesnes2004, solanki2010, ouakad2010}. NEMS is a natural advancement of microelectromechanical systems (MEMS). In NEMS, nanotubes and nanowires with circular cross-sections form a new class of beam structures while in MEMS rectangular cross-section beams are prevalent. Many efforts have been devoted to understand the characteristics of microbeams \cite{batra2007, joglekar20101, younis2003, joglekar20102}. \\
\indent Galerkin based reduced order modelling is an important technique which is widely used in MEMS analysis \cite{younis2003, batra2007}. In this method, the governing integro-partial differential equation (IPDE) of motion is reduced to a system of ordinary differential equations (ODEs). This is done by assuming the solution in terms of pre-assumed spatial functions (mainly linear mode shapes of a beam) and unknown time functions, and then substituting the assumed solution into the IPDE to obtain the required system of ODEs. In recent works, Ouakad and Younis \cite{ouakad2010} and Rasekh et al. \cite{rasekh2010} have carried out static and dynamic analysis of nanotube oscillators using reduced order model (ROM) technique. Rasekh et al. have used single-mode ROM for analysis while Ouakad and Younis have used multi-modal ROM, and observed that only single basis function (first linear mode) is sufficient to obtain a converged solution.\\
\indent Residual stress, often introduced during fabrication, alters the designed characteristics of actuators and oscillators \cite{pasquale2010}. Fixed-fixed type beams are dominated by axial thermo-mechanical residual stress, originating in the difference in thermal expansion coefficients of the beam and supporting structure. Residual stress has been considered by some researchers on rectangular cross-section microbeams. Abdel-Rahman et al. \cite{rahman2002} have solved static and eigenvalue problem of microbeams using shooting technique. They included the effect of residual stress in their study. In eigenvalue analysis, free vibration characteristics of a microbeam at deflected position due to applied DC voltage has been studied. Kuang and Chen \cite{kuang2004} and Jia et al. \cite{jia2010} have done eigenvalue analysis of microbeams, including the effect of residual stress, using differential quadrature method. Soma et al. \cite{soma2010} have studied experimentally and theoretically the effects of residual stress in pull-in behaviour of micro-switches. Pull-in is a well-known instability point beyond which elastic restoring force cannot balance applied electrostatic force and beam snaps down to the gate electrode \cite{joglekar20101}.\\
\indent  The focus of this study is a circular cross-section nanobeam with fixed-fixed end condition. The effects of both tensile and compressive type axial residual stresses have been studied using multi-modal ROM technique. The study comprises static deflection under applied DC voltage, pull-in parameters and variation of resonating frequency with applied DC voltage. The convergence of the ROM solution has also been studied. The results have been verified by numerically solving the boundary value problem using collocation technique. In addition, a mechanism is suggested for resonating frequency stability under temperature variation.
\section{Mathematical model}
\begin{figure}[h]
\begin{center}
\includegraphics[width=8cm]{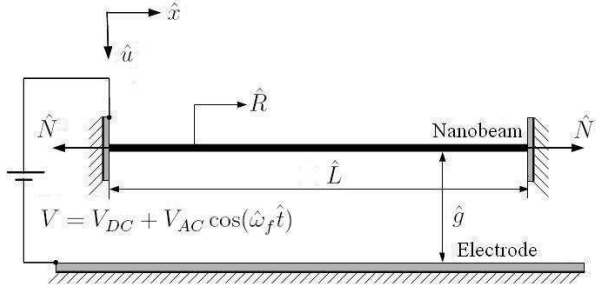}
\caption{Schematic diagram of an electrostatically actuated nanobeam}
\label{fig2.1}
\end{center}
\end{figure}

A schematic diagram of an electrostatically actuated cylindrical nanobeam is depicted in Fig. \ref{fig2.1}. The beam has clamped-clamped end condition with length $\hat L$ and radius $\hat R$. It is in capacitive arrangement with a electrode plate where $\hat g$ is the distance of separation between beam and plate. The hat represents physical quantities with dimension to differentiate from their non-dimensional forms. The beam is actuated by a DC load $\hat V_{DC}$ superimposed an AC harmonic load of amplitude $\hat V_{AC}$ and driving frequency $\hat \omega_f$. The beam is vibrated in viscous medium which has the damping coefficient $\hat c$. The residual stress is assumed uniform throughout the beam and is axial in nature. Here axial load $\hat N$ represents the residual stress of the beam. $\hat u(\hat x, \hat t)$ is vertical deflection of the nanobeam along spatial coordinate $\hat x$ at time $\hat t$.
 The equation of motion for the displacement $\hat u(\hat x, \hat t)$ is \cite{ouakad2010, rasekh2010}
\begin{equation}\label{eq2.2}
\begin{array}{l}
\displaystyle
 \hat E\hat I\frac{{{\partial ^4}\hat u}}{{\partial {{\hat x}^4}}} + \hat \rho \hat A\frac{{{\partial ^2}\hat u}}{{\partial {{\hat t}^2}}} + \hat c\frac{{\partial \hat u}}{{\partial \hat t}} = \left[ {\frac{{\hat E\hat A}}{{2\hat L}}\int\limits_0^L {{{\left( {\frac{{\partial \hat u}}{{\partial \hat x}}} \right)}^2}d\hat x + \hat N} } \right]\frac{{{\partial ^2}\hat u}}{{\partial {{\hat x}^2}}} \\
 \displaystyle
  + \frac{{\pi \varepsilon_0 {{\left( {{{ V}_{DC}} + {{ V}_{AC}}\cos (\hat \omega_f \hat t)} \right)}^2}}}{{\sqrt {{{\left( {\hat g + \hat R - \hat u} \right)}^2} - {{\hat R}^2}} {{\left[ {{{\cosh }^{ - 1}}\left( {\frac{{\hat g + \hat R - \hat u}}{{\hat R}}} \right)} \right]}^2}}} ;
 \end{array}
\end{equation}
and the boundary conditions are
\begin{equation*}\label{eq2.21}
{\left. {\frac{{\partial \hat u}}{{\partial \hat x}}} \right|_{\hat x = \,0}} = {\left. {\frac{{\partial \hat u}}{{\partial \hat x}}} \right|_{\hat x = \,1}} = {\left. {\hat u} \right|_{\hat x = \,0}} = {\left. {\hat u} \right|_{\hat x = \,1}} = 0 \cdot
\end{equation*}
$\hat A$ and $\hat I$ represent the cross-section area and the area moment of inertia of the nanobeam. $\hat E$ and $\hat \rho$ are Young's modulus and density of the nanobeam and $\varepsilon_0$ is the permittivity of vacuum. All terms in the left hand side of (\ref{eq2.2}) are same as in a equation of free vibrating linear beam. The first term of the right hand side of (\ref{eq2.2}) includes nonlinear stretching term due to deflection and residual stress whereas the second nonlinear term is actuation force due to applied DC and AC voltage.\\
\indent For convenience, (\ref{eq2.2}) is represented in terms of non-dimensional variables
\begin{equation*}\label{eq2.3}
u = \frac{{\hat u}}{{\hat g }},\,\,\,\,\,x = \frac{{\hat x}}{\hat L} \,\,\,\text{and}\,\,\,\hat t = \frac{{\hat t}}{\hat T},
\end{equation*}
where $ \hat T = \sqrt{\hat \rho \hat A \hat L^4 /\hat E \hat I}$ is a time constant. By substituting  the non-dimensional variables in (\ref{eq2.2}), the modified equation can be written as
\begin{equation}\label{eq2.4}
{u^{''''}} + \ddot u + c\dot u = \left[ {{\alpha _1}\int\limits_0^1 {{u^{{'^2}}}} dx + N} \right]{u^{''}} + {\alpha _2}{V^2}f\left( u \right),
\end{equation}
where
\begin{equation*}\label{eq2.87}
f(u) = \frac{1}{{\sqrt {{{\left( {1 + {R_0} - u} \right)}^2} - {R_0}^2} {{\left[ {{{\cosh }^{ - 1}}\left( {\frac{{1 + {R_0} - u}}{{{R_0}}}} \right)} \right]}^2}}},
\end{equation*}
\begin{equation*}\label{eq2.9}
c = \frac{{\hat c \hat L^4 }}{{\hat E \hat I \hat T}},\,\,\,\,\,\alpha _1  = \frac{{\hat A \hat g^2 }}{{2 \hat I}},\,\,\,\,\, N = \frac{{\hat N \hat L^2 }}{{\hat E \hat I}},\,\,\,\,\,\alpha _2  = \frac{{\pi \varepsilon_0 \hat L^4 }}{{\hat g^2 \hat E \hat I}},
\end{equation*}
\begin{equation*}
V={{V_{DC}} + {V_{AC}}\cos ( \omega_f t)},\,\,\,\omega_f= \hat \omega_f \hat T \,\,\text{and} \,\,\, R_0  = \frac{\hat R}{{\hat g}} \cdot
\end{equation*}
The overhead dot $^{.}$ and prime $^{'}$ represent partial derivatives with respect to time $t$ and spatial coordinate $x$ respectively. In the modified equation (\ref{eq2.4}), AC and DC voltage terms are kept in dimensional form. Static analysis of a nanobeam under the action of applied DC load $V_{DC}$ is presented in next section.
\section{Static analysis}
\begin{figure}[h]
\includegraphics[width=8cm]{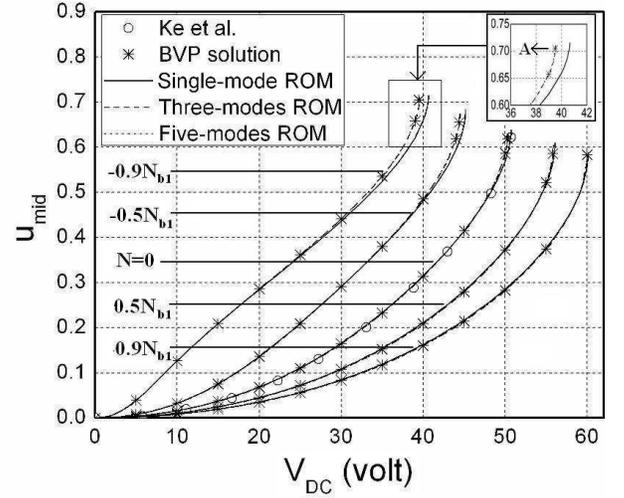}
\caption{Static normalised mid-point deflection $u_{mid}$ with DC voltage $V_{DC}$ for different values of residual stress; $N_{b1}$ is first Euler buckling load (Inset: magnified portion of near pull-in point curve of the case $N=0.90N_{b1}$)}
\label{fig4.1}
\end{figure}

\begin{figure}[h]
\includegraphics[width=8cm]{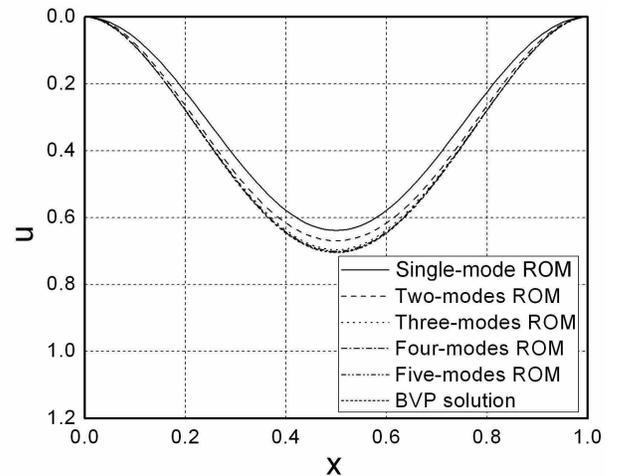}
\caption{Curvature of the deflected beam of the case $N=-0.90 N_{b1}$ at $V_{DC}=39.5 V$  }
\label{fig4.2}
\end{figure}

Static response of a nanobeam whose properties is given in the Table \ref{tab4.1} is studied. This nanobeam model has also been analysed by Ouakad and Younis \cite{ouakad2010} and Ke et al. \cite{ke20052}.
\begin{table}[here]
\caption{Properties of the nanobeam which is used for analysis}
\label{tab4.1}
\label{tab4.1}
\begin{center}
\begin{tabular}{|c|c|c|c|c|}
  \hline
 Length \emph{L} & Radius \emph{R} & Initial gap \emph{g}  & Young's\\
  (nm) & (nm) & (nm) & modulus \emph{E}(GPa) \\\hline
  3000 & 30 & 100 & 1000  \\
  \hline
\end{tabular}
\end{center}
\end{table}
The equation for the static analysis can be derived from (\ref{eq2.4}) by setting the time derivative terms and dynamic forcing term to zero. This ordinary differential equation falls in the class of two point boundary value problems. The problem has been solved numerically in MATLAB environment using ROM technique and the solution has been verified independently by solving the two point boundary value problem (BVP) by collocation method using $\mathtt{bvp4c}$ solver \cite{shampine2003}.\\
\indent For developing the ROM, the solution of (\ref{eq2.4}) can be assumed as \cite{younis2003, ouakad2010}
\begin{equation}\label{eq3.1}
    u(x) = \sum\limits_{i = 1}^m a_i{\phi _i (x)} \cdot
\end{equation}
The spatial function $\phi_i$ is $i^{th}$ linear undamped symmetric mode shape of a straight beam under axial load $N$ and $a_i$ is unknown scalar constant. The modes are orthogonal in nature and have been calculated analytically \cite{bokaian1988}. The modes were normalised such that $\int^1_0 \phi_i^2 dx = 1$. The ROM has been obtained by substituting the assumed solution (\ref{eq3.1}) of $u(x)$ in (\ref{eq2.4}), multiplying it by $\phi_n$ and integrating from $x=0$ to $x=1$. The time-independent form of ROM for static analysis is \\
\begin{equation}\label{eq3.8}
\begin{array}{l}
 \omega _n^2 a_n  = \left( {\alpha _1 \sum\limits_{i,\,j,\,k\, = 1}^m {a_i } a_j a_k \int\limits_0^1 {(\phi _i^{'} \phi _j^{'} } )dx\, \cdot \int\limits_0^1 {(\phi _k^{''} \phi _n )} dx} \right)
  \\+{\alpha _2}V_{DC}^2\int\limits_0^1 {f\left( {\sum\limits_{i = 1}^m {{a_i}{\phi_i}} } \right){\phi _n}} dx, \,\,\,\,\,\, \text{for \emph{n} = 1, 2...\emph{m}} \cdot\\
\end{array}
\end{equation}
\indent This is a system of nonlinear algebraic equation, where the number of equations in the system is equal to number of modes that has been used to develop the ROM. The problem has been solved with five ROM models (based on one to five modes). \\
\indent For construction of solution from BVP, the definite integral term (refer (\ref{eq2.4})) has been considered as a constant. The $\mathtt{bvp4c}$ solver requires a good initial guess of the solution since the solver searches the solution near the provided guess. Convergence problem was encountered during solving the BVP problem for larger applied DC voltage near pull-in. This issue has been resolved using the single-mode ROM solution as an initial guess. \\
\indent Figure \ref{fig4.1} shows the variation of normalised deflection of the mid-point $u_{mid}$ of the beam with DC voltage $V_{DC}$ for five different cases of axial load $N$: $-0.90N_{b1}$, $-0.50N_{b1}$, 0,  $0.50N_{b1}$,  $0.90N_{b1}$. Here $N_{b1}$ ($\hat N_{b1} = 4\pi^2 \hat E \hat I/ \hat L^2$ and $N_{b1} = 4 \pi^2$) is the first buckling load of a fixed-fixed Euler-Bernoulli beam in non dimensional form \cite{bokaian1988}. For each case of $N$, single-mode, three-modes and five-modes ROM results have been compared with the BVP solution. The ROM and BVP results have also been compared with the result of Ke et al. \cite{ke20052} for the case $N=0$. The results of current work are in good agreement with the published result. Ke et al. have obtained the solution using the finite difference method. The same nanobeam has also been analysed for the case $N=0$ by Ouakad and Younis \cite{ouakad2010} and they observed that single-mode ROM provides sufficiently converged result. This is also evident here for case $N=0$, however as can be seen from Fig. \ref{fig4.1} higher modes start to play an increasingly important role near the pull-in point when a high compressive stress exists in the beam. From Fig. \ref{fig4.1}, it can also be deduced that as residual stress increases, pull-in voltage increases and pull-in displacement decreases. Pull-in voltage for the case $N=0.90N_{b1}$ is 50\% more than the case $N=-0.90N_{b1}$ while pull-in displacement for the case $N=0.90N_{b1}$ is 18\% less than the case $N=-0.90N_{b1}$.\\
\indent The convergence of the solution has also been studied throughout the beam deflection while Fig. \ref{fig4.1} shows convergence of ROM solutions in term of $u_{mid}$. Fig. \ref{fig4.2} shows the static deflection curve of the beam for the case $N=-0.90N_{b1}$ near pull-in point at $V_{DC}=39.5$V. This point is also marked with letter A in the inset in Fig. \ref{fig4.1}. The error in the estimation of $u_{mid}$ using single-mode ROM at this point is around 8.5\%, while the error in the estimation of pull-in voltage for the case $N=-0.90N_{b1}$ is around 3\%. From Fig. \ref{fig4.2} and Fig. \ref{fig4.1}, it can be deduced that three-modes ROM is suitable for accurate analysis even near pull-in. In next section, analysis of free vibration of the nanobeam at static deflected position is presented.
\section{Free vibration analysis}
\begin{figure}[h]
\includegraphics[width=8cm]{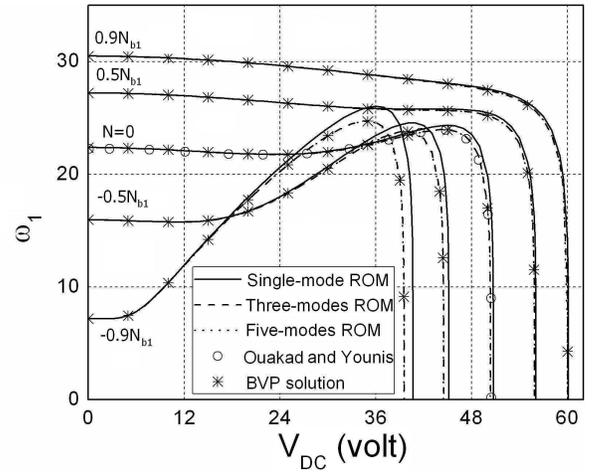}
\caption{Resonating frequency $\omega_1$ variation with applied DC voltage $V_{DC}$ for different values of residual stress; $N_{b1}$ is first Euler buckling load }
\label{fig5.1}
\end{figure}

\begin{figure}[h]
\includegraphics[width=8cm]{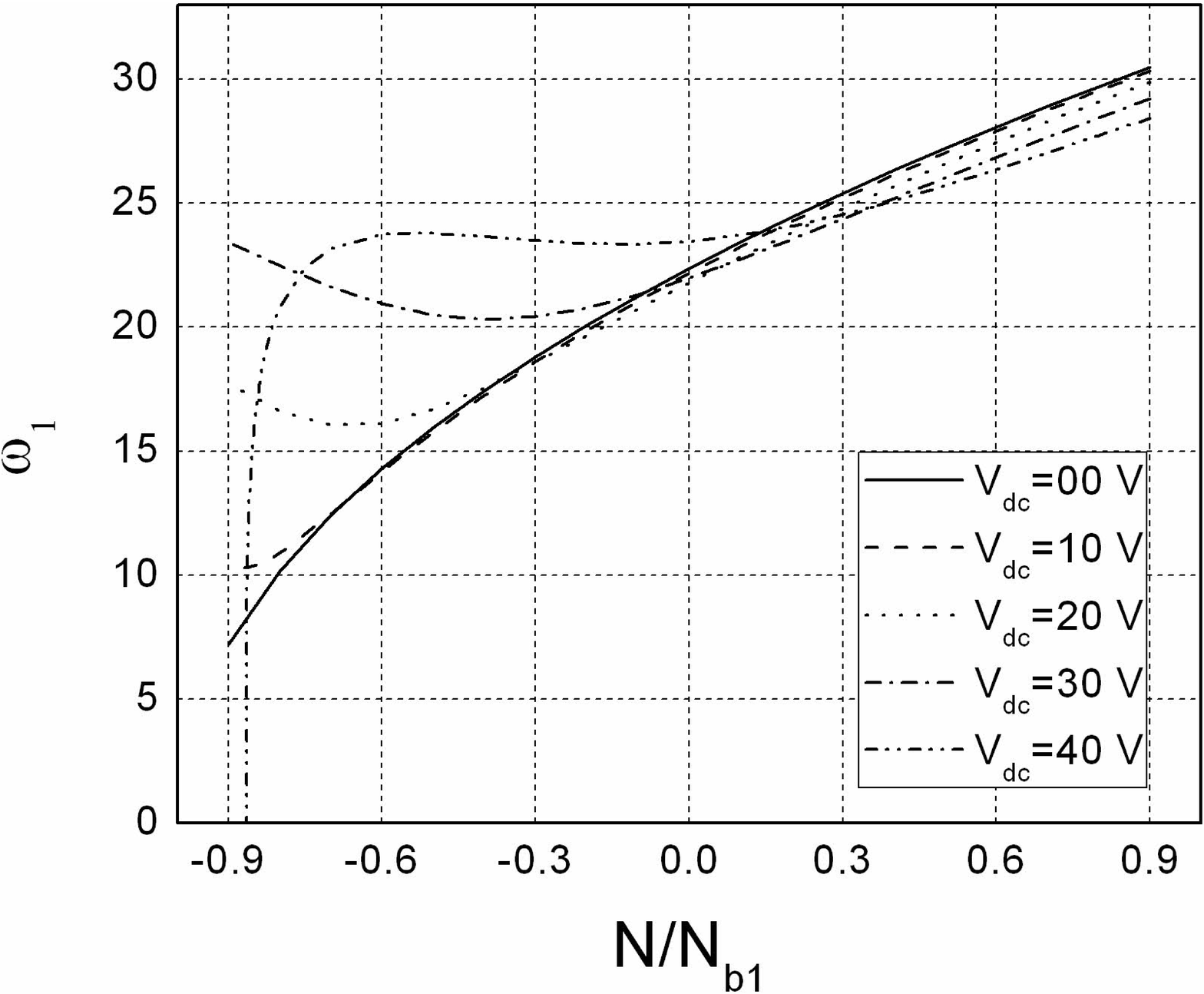}
\caption{Resonating frequency $\omega_1$ variation with normalised axial load $N/N_{b1}$ for different values of applied DC voltage $V_{DC}$}
\label{fig5.4}
\end{figure}

The nonlinearities in the dynamic model (\ref{eq2.2}) cause the resonating frequency of the nanobeam to vary with static deflection under action of applied DC voltage. A linear vibration problem at the deflected position is required to be solved to determine resonating frequency. For developing linear vibration model, the solution of (\ref{eq2.4}) can be assumed as
\begin{equation}\label{eq5.1}
u(x, t)=u_s(x)+u_d(x,t) \cdot
\end{equation}
 The electrostatic force due to applied DC voltage determines the static displacement $u_s(x)$, whereas  $u_d(x, t)$ is the linear vibration component. It is assumed that the $u_d (x, t)$ has very small magnitude with respect to static deflection $u_s(x)$ and the magnitude of $V_{AC}$ is also very small compared to $V_{DC}$. After substituting (\ref{eq5.1}) in (\ref{eq2.4}) and neglecting the terms containing higher order or combination of $u_d(x, t)$, derivatives of $u_d(x, t)$ and $V_{AC}$; a linear integro-partial-differential equation has been obtained for small vibration about the deflected position. The method of separation of variable has been applied by assuming the solution as $u_d(x, t) = u_{dt}(t)\phi_d(t)$ to obtain two uncoupled ordinary differential equation. The equation in spatial coordinate is the characteristic equation, which provides mode shapes and natural frequencies of free vibration about the deflected position. The characteristic equation is
\begin{equation}\label{eq5.3}
\begin{array}{l}
{\phi_d ^{''''}} - \left( {{\alpha _1}\int\limits_0^1 {{u_s}^{{{'}^2}}} dx + N} \right){\phi_d ^{''}} - \left( {2{\alpha _1}\int\limits_0^1 {{u_s}^{'}{\phi _d^{'}}} dx} \right){u_s}^{''} \\
- {\alpha _2}{V^2_{DC}}{f^{'}}({u_s})\phi_d  - {\omega _{d}}^2\phi_d  = 0\cdot
\end{array}
\end{equation}
Here $f^{'}$ is partial derivative of $f$ with respect to $u$. This is an eigenvalue problem whose solution gives mode shapes $\phi_d$ and frequencies $\omega_d$. This eigenvalue problem has been solved by ROM and collocation methods similar as was done in the static analysis.\\
\indent The solution of (\ref{eq5.3}) can be assumed as linear combination of modes of straight beam for developing reduced order model (ROM) as
\begin{equation}\label{eq5.31}
{\phi _d} = \sum\limits_{i = 1}^m {{b_i}{\phi _i}}.
\end{equation}
The ROM has been obtained by substituting assumed solution (\ref{eq5.31}) in (\ref{eq5.3}) then multiplying it by $\phi_n$ and integrating from $x=0$ to $x=1$; the resulting ROM is obtained as
\begin{equation}\label{eq5.32}
\begin{array}{l}
 {b_n}\omega _n^2 - \sum\limits_{i = 1}^m {{b_i}} \left( {{\alpha _2}V_{DC}^2\int\limits_0^1 {{f^{'}}\left( {{u_s}} \right){\phi _i}{\phi _n}dx} } \right) - \\
  \sum\limits_{i = 1}^m {{b_i}\left( {{\alpha _1}\int\limits_0^1 {u_s^{{{'}^2}}dx}  \cdot \int\limits_0^1 {\phi _i^{''}{\phi _n}dx + 2{\alpha _1}} \int\limits_0^1 {u_s^{'}\phi _i^{'}dx \cdot } \int\limits_0^1 {u_s^{''}{\phi _n}dx} } \right) }  \\
 = {b_n}\omega _d^2 \,\,\,\,  \text{for \emph{n} = 1, 2... \emph{m}} \cdot
  \end{array}
\end{equation}
This is a linear algebraic eigenvalue problem. The natural frequencies are square roots of eigenvalues and the mode shapes at deflected position can be calculated from eigenvectors. The eigenvalue problem (\ref{eq5.3}) has also been solved by collocation technique for verifying the results. Mode shapes of a straight beam have been used as initial guesses in $\mathtt{bvp4c}$ solver.\\
\indent For different values of residual stress, Fig. \ref{fig5.1} shows variation of first resonating frequency $\omega_1$ with applied DC voltage $V_{DC}$. For $N=0$, the results of current work has been compared with Ouakad and Younis \cite{ouakad2010} result. There is a good agreement between results of current work and the published result. The observation of static analysis also holds here that the converged ROM solution can be obtained from three-modes ROM, and higher modes start to contribute in the solution near pull-in point when large magnitude of compressive residual stress is present in the beam. For the case $N=-0.90N_{b1}$ at $V_{DC}=36$V, the error in estimation of $\omega_1$ from single-mode ROM is around 6\%. \\
\indent The variation of $\omega_1$ versus $V_{DC}$ is exploited for electrostatic tuning to obtain desired resonating frequency during operation of a NEMS resonator \cite{peng2010, solanki2010}. Fig. \ref{fig5.1} shows that presence of compressive residual stress significantly increases the tunability of the nanobeam resonator. In the case $N=-0.90N_{b1}$, resonating frequency $\omega_1$ is 7.19 at $V_{DC}=0$ and it increases to  24.75 at $V_{DC} = 35 V$. Here $\omega_1$ can be increased to more than three times.\\
\indent Variation of resonating frequency with residual stress has also been studied. Figure \ref{fig5.4} shows this variation for different cases of applied DC voltage. The results are obtained with three-modes ROM method. The resonating frequency $\omega_1$ shows very high sensitivity on residual stress in absence of applied $V_{DC}$ but sensitivity decreases under action of $V_{DC}$. Many resonator devices have a requirement of minimal resonating frequency variation in practical operating range of temperature \cite{hopcroft2004, melamud2007}. One main reason for the shift of resonating frequency under temperature variation is build up of internal stresses due to different thermal expansion coeeficients of beam and substrate \cite{hopcroft2004}. The internal stress is similar in nature as residual stress. The nanobeam resonator under investigation shows excellent frequency stability or very low sensitivity to internal stress in the case $V_{DC}$= 40 V. Variation of $\omega_1$ is around 1.5\% in the window of internal axial load $\pm 0.10 N_{b1}$; this is six times better than the case $V_{DC}=0$. In summary, residual stress and biased voltage $V_{DC}$ significantly vary the resonating frequency of a nanobeam resonator and interplay between them may be utilised to increase the resonating frequency tunability (Fig. \ref{fig5.1}) and to achieve frequency stability under temperature variation (Fig. \ref{fig5.4}). The next section concludes the work presented in this paper.
\section{conclusion}
Reduced order modelling is a simple method for analysis of electrostatically actuated nanotubes and nanowires based devices. In this work, a cylindrical nanobeam is analysed to study its static and dynamic characteristics. Single-mode reduced order model (ROM) is capable of providing accurate results when the nanobeam is free from compressive residual stress. However the analysis shows that at least three symmetric modes in ROM is necessary for accurate analysis near pull-in instability, when high magnitude of compressive residual stress is present.\\
\indent Residual stress has significant effects on the pull-in parameters and resonating frequency tunability characteristics. Presence of compressive residual stress decreases the pull-in voltage and increases the pull-in displacement. And tunability of resonating frequency improves significantly in presence of high magnitude of compressive residual stress.\\
\indent An inherent resonating frequency stability mechanism under temperature variation in nanobeam resonators is explored in this investigation. Variation in operating temperature causes internal stress and it is similar in nature to the residual stress. By studying the variation of resonating frequency with internal stress under action of constant DC voltage, it has been found that the resonating frequency sensitivity with internal stress decreases significantly at a particular DC voltage. This is an important property which may be considered during design of nanobeam resonators.\\
\section*{Acknowledgements}
The authors gratefully acknowledge financial support from the Industrial Research and Consultancy Centre, India Institute of Technology Bombay and the Department of Science and Technology, Government of India. We are also thankful to Mandar M. Desmukh and Hari S. Solanki of the Tata Institute of Fundamental Research, India for useful discussions.
\bibliography{nanobeam}
\end{document}